\documentstyle[twoside,fleqn,espcrc2]{article}

\newtheorem{Theorem}{Theorem}

\newcommand{\nn}{\nonumber}
\newcommand{\non}{\nonumber\\}
\newcommand{\be}{\begin{equation}}
\newcommand{\ee}{\end{equation}}
\newcommand{\ben}{\begin{displaymath}}
\newcommand{\een}{\end{displaymath}}
\newcommand{\baa}{\begin{eqnarray}}
\newcommand{\eaa}{\end{eqnarray}}

\newcommand{\z}{\zeta}
\newcommand{\D}{\Delta}
\renewcommand{\d}{\delta}

\newcommand{\m}{\mu}

\def\c{\cite}
\def\a{\alpha}
\def\b{\beta}
\def\g{\gamma}
\def\d{\delta}

\def\C{{\bf C}}

\def\bi{\bibitem}

\def\tr{{\rm tr}}
\def\i{\infty}
\def\E{{\cal E}}

\def\cO{{\cal O}}

\def\x{\xi}
\def\xb{{\bar{\xi}}}
\def\zb{{\bar z}}
\def\be{\begin{equation}}
\def\ee{\end{equation}}
\def\la{\label}
\def\c{\cite}
\def\f{\frac}
\def\Eb{\bar{\cal E}}
\def\ba{\begin{array}}
\def\ea{\end{array}}
\def\r{\ref}
\def\Ref#1{(\ref{#1})}
\def\I{{\bf 1}}

\def\g{\gamma}
\def\R{{\bf R}}
\def\cC{{\cal C}}
\def\Cx{{\cC}^{(\x)}}
\def\Cxb{{\cC}^{(\xb)}}

\def\nn{\nonumber}

\def\s{\sigma}

\def\H{{\cal H}}
\def\p{\partial}
\def\X{{\cal X}}
\def\Y{{\cal Y}}

\def\cos{SL(2,\R)/SO(2)}
\def\zb{\bar{z}}
\def\gb{\bar{\g}}

\title{On $2D$ quantum gravity coupled to a $\s$-model
\thanks{Talk given by H.~Nicolai at the 29th International Symposium 
Ahrenshoop, Buckow, Germany, August 29 - September 2, 1995.}}

\author{D.\,Korotkin\thanks{On leave of absence from Steklov
Mathematical Institute, Fontanka, 27, St.Petersburg, 191011 Russia.},
H.\,Nicolai, H.\,Samtleben\bigskip\\II. Institut f\"ur Theoretische
Physik, Universit\"at Hamburg, Luruper Chaussee 149, D-22761 Hamburg,
Germany}
\begin{document}

\maketitle

\section{Introduction}
This contribution is a review of the method of isomonodromic
quantization of dimensionally reduced gravity developed in
\c{KN1,KN2,KN3}.  Our approach is based on the complete separation of
variables in the isomonodromic sector of the model and the related
``two-time" Hamiltonian structure. This allows an exact quantization
in the spirit of the scheme developed in the framework of integrable
systems \cite{Fad}. Possible ways to identify a quantum state
corresponding to the Kerr black hole are discussed. In addition, we
briefly describe the relation of this model with Chern Simons theory.

\section{The model}

The Lagrangian of $2D$ gravity coupled to a $\cos$ $\s$-model is
\be
{\cal L} =\rho \Big( h R + {\rm tr} (g_{z} g^{-1}
g_{\zb}g^{-1}) \Big),
\la{L}
\ee
where the metric has been brought into conformal gauge
\be
ds^2=h(z,\zb)d z d\zb ;
\la{metric}
\ee
$R=(\log h)_{z\zb}/h$ is the Gaussian curvature of the worldsheet,
$g\in \cos$ and $\rho\in \R$ is the dilaton.  The equation of motion
for $\rho$ derived from \Ref{L}
\be
\rho_{z\zb} = 0      \la{dilequ}
\ee
is solved by 
\be
\rho (z,\zb) = {\rm Im} \, \x( z),
\la{gaugefix}\ee
where $\x(z)$ is a (locally) holomorphic function.  Now we can further
specialize the gauge by identifying $\x$ with the worldsheet
coordinate.  Then the equation of motion for $g$ coming from \Ref{L}
is
\be
\big((\x-\xb)g_{\x}g^{-1}\big)_{\xb} + 
\big((\x-\xb)g_{\xb}g^{-1}\big)_{\x} =0 .
\la{ee}\ee
Using the following parameterization of an arbitrary $\cos$-valued
matrix:
\be
{g}=\f{1}{\E+\Eb}
 \pmatrix{ 2 & i(\E-\Eb)\cr
            i(\E-\Eb) & 2\E\Eb  \cr}
\la{g}\ee
in terms of the complex-valued function $\E(\x,\xb)$,
we can rewrite \Ref{ee} in the familiar form of the Ernst equation 
\cite{Ernst}:
\be
(\E +\Eb)\Big(\E_{\x\xb}-\f{\E_\x-\E_\xb}{2(\x-\xb)}\Big)=
2\E_\x \E_\xb
\la{eee}\ee
To get from \Ref{L} the remaining equations of motion for the
conformal factor $h$, we have to temporarily relax the conformal gauge
and to vary \Ref{L} with respect to the off-diagonal
elements of the metric. Restoring the conformal gauge then yields
\be
(\log h)_{\x} =\f{\x-\xb}{4}\tr(g_{\x} g^{-1})^2 \quad{\rm and~c.c.}
\la{h}
\ee

It is well-known \c{Exact} that the same equations of motion arise in
stationary axisymmetric reduction of the $4D$ Einstein equations.
The quantum theory based on \Ref{L} may therefore be regarded as an
example of the ``midi-superspace" approximation to $4D$ quantum gravity.

\section{Deformation equations and $\tau$-function}

Consider the following system of differential equations for $2\times 2$
matrices $\{A_j (\x,\xb)\}$ and the 
functions $\{\g_j(\x,\xb)\}$ with $j=1,...,N$:
\baa
\f{\p A_j }{ \p \x }&=&
\f{2}{\x-\xb}\sum_{k\neq j}\f{[A_k,\;A_j]}{(1-\g_k)(1-\g_j)}\la{1}\\
\f{\p A_j }{ \p \xb }&=&
\f{2}{\xb-\x}\sum_{k\neq j}\f{[A_k,\;A_j]}{(1+\g_k)(1+\g_j)}
\nn
\eaa
\baa
\f{\p \g_j}{\p\x}&=&\f{\g_j}{\x-\xb}\f{1+\g_j}{1-\g_j} \la{pe}\\
\f{\p\g_j}{\p\xb}&=&\f{\g_j}{\xb-\x}\f{1-\g_j}{1+\g_j} 
\nn
\eaa
These (compatible) equations are solved by
\baa
\g_j&=&\f{2}{\x-\xb}\times \la{gamma}\\
&& \left\{w_j -\f{\x+\xb}{2} 
\pm\sqrt{(w_j-\x)(w_j-\xb)}\right\},\nn
\eaa
where $w_j\in\C$ are constants of integration; in the sequel we
shall assume $\g_j$ to be defined by \Ref{gamma}. One can easily
check that the system \Ref{1} is always compatible if \Ref{pe} holds.

Next define the $\tau$-function $\tau(\x,\xb)$ associated with
\Ref{1} by
\be
d\log\tau =\sum_{j< k} \tr (A_j A_k)d\log(\g_j-\g_k),
\la{tau}\ee
where the differential is to be taken with respect to the variables
$(\x,\xb)$. Equivalently it can be computed with respect to the
variables $\{\g_j\}$, which gives $\tau$ as a function of the 
parameters $\g_j$. Notice that the $1$-form on the r.h.s. of
\Ref{tau} is always closed as a consequence of \Ref{1}.

It is easy to check that $\tr A_j$, $\tr A_j^2$ and
$\sum_{j=1}^N A_j$ are integrals of motion of the system \Ref{1}.

Our purpose will be to exhibit the link between the system \Ref{1}
with $\g_j$ given by \Ref{gamma} and the equations of motion \Ref{ee}
and \Ref{h} of the previous section. A partial answer is given by

\begin{Theorem}
Let $\{A_j\}$ be an arbitrary solution of the system \Ref{1}
with $\g_j$ given by \Ref{gamma}. Then 
\begin{enumerate}
\item
The system of equations 
\baa
g_{\x} g^{-1}=\f{2}{\x-\xb}\sum_{j}\f{A_j}{1-\g_j}\la{cur}\\ 
g_{\xb} g^{-1}=\f{2}{\xb-\x}\sum_{j}\f{A_j}{1+\g_j}\nn
\eaa
for the matrix-valued function $g(\x,\xb)\in GL(2,\C)$  is always 
compatible.
\item
The solution $g(\x,\xb)$ of this system satisfies  equation \Ref{ee}. 
\item
The conformal factor $h$ defined by \Ref{h} is related to the
$\tau$-function of the system \Ref{1} as follows:
\be
h =
C(\x-\xb)^{\f{1}{2} \tr A_\i}   \prod_j
\left\{\f{\partial \g_j}{\partial w_j}\right\}^{\f{1}{2}\tr A_j^2}
\tau 
\la{link}\ee  
where $A_\i\equiv \sum_{j=1}^N A_j$ and $C$ is a constant.
\end{enumerate}
\la{main}
\end{Theorem}
To understand the precise correspondence between the solutions of \Ref{1}
and the original model, one has to ensure the coset and reality
conditions $g\in \cos$ and $h\in\R$.
To this aim we define the rational function $A(\g)$ by
\be
A(\g)=\sum_{j=1}^N \f{A_j}{\g-\g_j} .
\la{Ag}\ee
The proof of the following theorem may be found in \c{KN3}.

\begin{Theorem}
Let $\{A_j\}$ be a solution of the system \Ref{1} satisfying the following
additional conditions:
\begin{enumerate}
\item
Reality:
\be
\overline{A(\g )} = - A(-\overline{\g})  \la{Areality}
\ee
\item
Asymptotic regularity:
\be
A_\i\equiv\sum_{j=1}^N A_j =0
\la{reginf}\ee
\item
Invariance of the $\tau$-function with respect to the involution
$\g_j \rightarrow 1/\g_j$:
\be
\tau\Big(\f{1}{\g_j},\dots,\f{1}{\g_j}\Big)=
c_0 \tau(\g_1,\dots,\g_N)
\la{sym}\ee
with some constant $c_0 \neq 0$.
\end{enumerate}
Then the constants of integration in \Ref{h} and \Ref{cur} may be
chosen in such a way that $h\in \R$ and $g\in\cos$.
\la{constr}
\end{Theorem}
At this point the relation between \Ref{1} and
the original model may still appear obscure;
it will be clarified in section \ref{csls}. Let us just emphasize
that the variables in the system \Ref{1} have been 
completely separated; thus we can treat the ``left" ($\x$) and ``right"
($\xb$) moving sectors as completely independent.

The link between the system \Ref{1} and the classical
Schlesinger equations \c{Sch} for the variables $A_j$ considered as
functions of $\g_1,...,\g_N$ 
\baa
\f{\p A_{j}}{\p\g_k}
            &=&\f{[A_j,A_k]}{\g_j-\g_k}\;\;\;\;(k\neq j) 
\la{Sch}\\
\f{\p A_{j}}{\p\g_j}
            &=& - \sum_{i\neq j}
\f{[A_j,A_i]}{\g_j-\g_i}\nn
\eaa
is given by
\begin{Theorem}
Let $A_j(\{\g_j\})$ be a solution of the Schlesinger equations \Ref{Sch}
satisfying the constraint \Ref{reginf}. Then, assuming that all $\g_j$
depend on $(\x,\xb)$ according to \Ref{gamma}, the functions $A_j
(\x,\xb)$ solve system \Ref{1}.
\end{Theorem}

\section{An example: the Kerr-NUT solution}

The general solution of the system \Ref{1} for arbitrary values of $N$
and the parameters $w_j$ is certainly not possible.
However, one can try to understand which solutions of \Ref{1}
correspond to known solutions of \Ref{eee}.
For example, the Kerr-NUT solution of \Ref{eee} corresponds to
$N=4$,
\be 
w_1=w_3=-\s \;\;\;\;
w_2=w_4=\s\;\;\;\; \s\in \R
\la{wKerr}\ee
with $\g_3=\g_1^{-1}$ and
$\g_4=\g_2^{-1}$.  The integrals of motion $\tr A_j^2$ should equal
$1/2$ (since $\tr A_j=0$, this means that the eigenvalues of $A_j$ are
equal to $\pm 1/2$). It is not difficult to show that the solution
$\{A_j\}$ of \Ref{1} satisfying these conditions and the constraints
given by Thm.\ref{constr} corresponds to the Ernst potential
\be
\E=\f{(\b_2-\b_1)\X - (\b_2+\b_1)\Y -2}
{(\b_2-\b_1)\X-(\b_2+\b_1)\Y +2},
\la{Kerr}\ee
where
\be
\X=\f{1}{2\s}\{S_1 +S_2\}\;\;\;\;\;
 \Y=\f{1}{2\s}\{S_1 -S_2\},
\la{XY}\ee
with
\baa
S_1&=& \sqrt{(\x+\s)(\xb+\s)} \non
S_2 &=&\sqrt{(\x-\s)(\xb-\s)} \nn
\eaa
are prolate ellipsoidal coordinates; $\b_{1,2}$ are complex
constants satisfying $|\b_1|=|\b_2|=1$.  This is nothing but the
Kerr-NUT solution; the Kerr solution itself corresponds to
$\b_2=-\b_1$.

\section{Two-time Hamiltonian structure}

We adopt here a ``two-time" Hamiltonian formalism with the two
``times" corresponding to the lightcone coordinates $\x$ and
$\xb$. One major advantage of this procedure is that the quantum
theory is manifestly covariant under $2D$ coordinate transformations,
a feature which is far from obvious (and possibly not even true) for
the ADM formulation of canonical quantum gravity (see
e.g. \c{multi-time} for a recent discussion). Moreover, we must to
treat the ``times" $\x$ and $\xb$ as phase space variables because
they are really fields in a special gauge; then,
according to the general canonical procedure, the related total
Hamiltonians should weakly vanish, i.e. should be considered as
first-class constraints.  

The Hamiltonian structure which gives the complete set of equations of
motion in terms of the variables $\{A_j\}$, $\x,\xb$, $(\log h)_\x$
and $(\log h)_\xb$ is described by the following

\begin{Theorem}
The system \Ref{h}, (\r{1}) is a ``two-time"
Hamiltonian system with respect to the Poisson brackets
\be
 \left\{A(\g)  \stackrel{\otimes}{,}  A(\mu)\right\} =
\Big[ r\, ,\, A(\g)\otimes \I + \I \otimes A(\mu)\Big]
\la{pb}
\ee
\baa
\{ \x, (\log h)_\x \} &=& \{ \xb, (\log h)_\xb  \} ~=~1 
\la{pbh}\\
\{ \xb,  (\log h)_\x  \} &=& \{ \x, (\log h)_\xb \} ~=~0
\non
\{A_j, (\log h)_\x\}&=&\{A_j, (\log h)_\xb\} ~=~0,
\la{pbhA}
\eaa
where $A(\g)$ is given by \Ref{Ag} and the classical rational
$R$-matrix $r$ is equal to 
\ben
r(\g-\mu)=\frac1{\g-\mu} \left(\ba{cccc}1\;\;0\;\;0\;\;0\\
                                        0\;\;0\;\;1\;\;0\\
                                        0\;\;1\;\;0\;\;0\\
                                        0\;\;0\;\;0\;\;1\ea\right) 
\een
The mutually commuting Hamiltonian constraints 
in the $\x$ and $\xb$-directions are given by
\begin{eqnarray}
\Cx &:= &-(\log h)_\x + \f{1}{\x-\xb} {\rm tr} \, A^2 (1)
\nonumber \\
\Cxb & :=&-(\log h)_\xb +  \f{1}{\xb-\x}{\rm tr} \, A^2 (-1)
 \la{Ham}
\end{eqnarray}
\la{Poisson}
\end{Theorem}

This theorem can be verified by direct calculation.  The weak
vanishing of $\Cx$ and $\Cxb$ implies the equations \Ref{h} relating the
gravitational and matter degrees of freedom.  Commutativity of the
Hamiltonian constraints may be obtained by use of the general relation
\be \Big\{ \tr A^2(\g)  \, , \, \tr A^2(\mu) \Big\} = 0,
\la{commute} \ee
which is valid for arbitrary $\g$ and $\mu$.  The commutativity of the
flows generated by $\Cx$ and $\Cxb$ is equivalent to the decoupling of
the classical equations of motion in \Ref{h} and \Ref{1} and may be
viewed as a direct consequence of the compatibility of the system
(\r{h}), \Ref{1}. In terms of the standard ``one time" canonical
formalism with $\rho$ as Euclidean time, the combination 
${\cal C}^{(\rho)}= 1/2i (\Cx-\Cxb)$ corresponds to the Hamiltonian or
Wheeler-DeWitt constraint while $1/2 (\Cx+\Cxb)$ corresponds
to the diffeomorphism constraint.

The ``time evolutions" of any functional $F$ are generated as usual by
commutation with the total Hamiltonian constraints $\Cx$ and $\Cxb$,
i.e.
\be
\f{d F}{d\x} = \{ \Cx , F \}  \;\;\; \;\;\; 
\f{d F}{d\xb} = \{\Cxb , F \}.  \la{evolve}
\ee
On the l.h.s. here we have the total derivatives with respect to
$\x$, $\xb$; 
the first term of $\Cx$ or $\Cxb$ generates the partial derivatives
with respect to the coordinates and the second term takes care of the
$(\x,\xb)$-dependence of $A_j$.
Observe that we have $(\Cx )^\dagger = \Cxb$. 

Defining
\be
A_{_j,\a \b} =: A_j^a t^a_{\a\b}, \la{Aab} \ee
where $t^a$ are the generators of $SL(2,\R)$, and inserting \Ref{Ag}
into \Ref{pb}, we get
\ben
\{A_j^a,\;A_k^b\}=2\delta_{jk}{f^{ab}}_c A_j^c ,
\een
where ${f^{ab}}_c$ are the structure constants of $SL(2,\R)$.

Observables in the sense of Dirac are by definition all 
those functionals $\cO$ on phase space which weakly 
commute with the constraints $\Cx$ and $\Cxb$,
but do not vanish on the constraint hypersurface 
$\Cx = \Cxb = 0$, i.e.
\be
\{ \Cx , \cO \} \approx 0 \;\;\; , \;\;\; \{ \Cxb , \cO \} \approx 0. 
\la{observable} \ee
By \Ref{evolve} the observables are independent of the coordinates and
therefore highly non-local objects as one would expect on general
grounds \c{QG,Ashtekar}.  First of all, the parameters $w_1,...,w_N$
trivially belong to this class since they commute with
everything. Second, and more importantly, the mono\-dromies
$M_1,...,M_N$ of the connection $A(\g) d\g$ defined by
\be
M_j={\cal P}\exp\oint_{l_j} A(\g) d\g,
\la{mon}\ee
where the contour $l_j$ starts at $\g=\i$ and encircles the point
$\g_j$, are also observables for arbitrary $N$. For a discussion of
this fact, see \c{KN3}.  All observables can be generated from the set
\be
Obs := \big\{w_1,...,w_N; M_1,...,M_N \big\}
\la{sbo}\ee
by taking products and linear combinations. In this sense,
$Obs$ constitutes a complete set of classical (and quantum) observables
for arbitrary $N$. These are the conserved ``non-local charges"
of dimensionally reduced gravity.

Notice also that the constraints mentioned in Thm.\ref{constr} are in
fact first class constraints with respect to our Poisson structure. 
In particular, the constraint $A_\i =0$ which closes into
the $SL(2,\R)$ algebra is nothing but the conserved charge which
generates the Ehlers transformations $g\rightarrow Q^t g Q$
with a constant matrix $Q\in SL(2,\R)$.

\section{Link to Chern Simons theory and the linear system}\label{csls}
It is known that the Ernst equation can be obtained as the compatibility
of a linear system \c{M,BZ}. The interpretation of the linear system 
as a zero curvature condition suggests a link with Chern Simons
theory whose equations of motion also state the vanishing of some
curvature. The new feature here is that the Chern Simons gauge 
connection lives on a space locally parameterized simultaneously
by the spectral parameter and the true space time coordinate.

The relevant Chern Simons action  (at level 1) reads
\be
S = \frac{1}{4\pi i} \int {\rm tr}\big(-A\partial_\xi A + 2A^\xi F\big) 
d\xi 
\label{action},
\ee
where $\xi$ plays the role of time, $A=A^\g d\g +
A^{\bar{\g}}d\bar{\g}$ is a time dependent connection 1-form
on the Riemann surface locally coordinatized by $\g, \bar{\g}$,
and $F\equiv F^{\g \bar{\g}} d\g d\bar{\g}$ is the curvature 2-form. 
The time component $A^\xi$ appears as a Lagrangian multiplier for the
first class constraints of vanishing curvature $F=0$:
\be
\{F^a(\g),F^b(\m)\} = 2\pi i f^{abc}F^c(\g)\d^{(2)}(\g-\m). \label{con}
\ee

In the usual treatment $A^\xi$ is gauged to zero which
leads to static components $A^\g$ and $A^{\bar{\g}}$. In particular,
the singularities of this connection are then time-independent and
treated by inserting static Wilson-lines in the action
(\ref{action}). Alternatively, we consider the gauge
\be
A^\xi(\g) = 
\frac{2A^\g(1)-\g(1+\g)A^\g(\g)}
{(\xi-\bar{\xi})(1-\g)}
\la{gc1}.
\ee
The residual gauge freedom corresponding to (\ref{con}) is fixed by 
demanding 
\be
A^{\bar{\g}}=0 \label{fix}
\ee
on the whole surface except for some set of zero measure. Because of 
(\ref{con}) and $F=0$ the remaining component $A^\g$ then becomes 
holomorphic
up to poles. To allow such singularities in $A^\g$ as in the previous
section, it is clear that (\ref{fix}) cannot be imposed everywhere
because the singularities arising via the relation
$\partial_{\bar{\g}}\frac1{\g} = 2\pi i\d^{(2)}(\g)$ 
would spoil the constraint (\ref{con}). Instead 
one should think of $A^{\bar{\g}}$ as being localized on some string
with endpoints at the singularities of $A^\g$.

The remaining equation of motion is
\be
\f{\p A^\x}{\p\g}-\f{\p A^\g}{\p\x}+[A^\x, A^\g] =0
\la{eqmot}\ee

The constraints can now be treated by introducing Dirac brackets.
The original Poisson-bracket that comes from the action (\ref{action}):
\be
\{A^{\g a}(\g),A^{\bar{\g} b}(\m)\} = 2\pi i  \d^{ab}\d^{(2)}(\g-\m)
\ee
is thereby changed to a bracket between the remaining meromorphic 
components $A^a(\g)\equiv A^{\g a}(\g)$ \cite{FR,KS}:
\be
\{A^a(\g),A^b(\m)\} = -f^{abc}\frac{A^c(\g)-A^c(\m)}{\g-\m}
\ee

This may be translated into a bracket structure on the coefficients of 
the poles of $A^{\g}$, which --- together with the positions of the 
poles --- now parameterize the phase space:
\be
\{A_i^a,A_j^b\} = 2\d_{ij}f^{abc}A_j
\ee
for 
\ben
A(\g) = \sum_{j}\frac{A_{j}}{\g-\g_j}
\een
It coincides with the Poisson structure introduced in
Thm. \ref{Poisson} of the previous section. The equations of motion
\Ref{eqmot} give rise to equations \Ref{1} and \Ref{pe}.
 
Among the surviving first class constraints is the total sum of the 
residues:
\ben
\int F(\g)  = A_{\infty} \approx 0
\een
as well as the Chern Simons Hamiltonian

\be
\Cx = \frac1{2\pi i} \int {\rm tr} A^\xi F, \label{CSH}
\ee  
which generates the equations of motion for the holomorphic 
component of the connection:
\ben
\partial_\xi A(\g) = \partial_\g A^\xi(\g) + [A^\xi(\g),A(\g)] 
\een

Splitting the Hamiltonian \Ref{CSH} it is now
possible to identify its parts with the expressions obtained in the
previous section.  A short calculation reveals
\ben
\frac1{2\pi i} {\rm tr} \int A^\xi \partial_{\bar{\g}}A^{\g} 
d\g d\bar{\g} = \frac1{\xi-\bar{\xi}} {\rm tr} A(1)A(1),
\een
such that defining
\ben
(\log h)_\xi \equiv \frac1{2\pi i} {\rm tr} 
\int A^\xi \Big(\partial_{\g}A^{\bar{\g}} + 
[A^{\bar{\g}},A^{\g}]\Big) d\g d\bar{\g},
\een
we have
\be
\{(\log h)_\x, A^\g\} = \f{\p A^\x}{\p\g}
\la{pbk}\ee
in agreement with \Ref{pbhA} if $A^\g$ is given by \Ref{Ag}.
All equations of motion are now generated by

\be
\Cx = - (\log h)_\xi +\f{1}{\x-\xb}\tr A^2 (1)
\ee

In this way the Poisson structure as well as the Hamiltonian and the
constraints have a natural explanation in the context of Chern Simons
theory. Similar considerations lead to the analogous results for the
$\bar{\xi}$-sector. However, further work is required to embed this
two-time treatment in one unified canonical approach.

It is quite instructive to see how the well-known auxiliary linear
system \c{M,BZ} arises in this framework. The analogous treatment of 
Chern Simons model in $(\g,\gb,\xb)$ space with the gauge choice
\be
A^\xb(\g) = 
\frac{2A^\g(-1)+\g(1-\g)A^\g(\g)}
{(\xi-\bar{\xi})(1+\g)}
\la{gc2}
\ee
gives the equation of motion supplementing \Ref{eqmot}:
\be
\f{\p A^\xb}{\p\g}-\f{\p A^\g}{\p\xb}+[A^\xb, A^\g] =0
\la{eqmot1}\ee
 
The vanishing of the curvatures \Ref{eqmot} and \Ref{eqmot1} implies the 
existence of a gauge transformation $\Psi(\g;\x,\xb)$ such that
\baa
\f{\p\Psi}{\p\g}= A^\g \Psi \;\;\;\;\; 
\f{\p\Psi}{\p\x}= A^\x \Psi \;\;\;\;\; 
\f{\p\Psi}{\p\xb}= A^\xb \Psi
\la{ls}\eaa
Substituting \Ref{gc1} and \Ref{gc2} into the last two equations and
using \Ref{cur} we get just the linear system of \c{BZ} with
$\g$ playing the role of the spectral parameter.  The solutions of
\Ref{ee} for which $A^\g$ can be represented as in \Ref{Ag}
are called isomonodromic; in particular, they contain all known
solutions such as multisoliton solutions \c{BZ} and the
algebro geometrical solutions of \c{Kor}, as well as many others.
Of course in assuming \Ref{Ag} we truncate the total phase space of the 
original model. We would expect that there exists a topology on the
space of solutions for which the isomonodromic solutions constitute
a dense subset of the phase space of ``all solutions"
(notice that the Poisson structure given by \Ref{pb},
\Ref{pbh} and \Ref{pbk} is independent of the ansatz \Ref{Ag}).

\section{Quantization}
To quantize the model, we replace the Poisson
brackets (\r{pb}) by commutators in the usual fashion:
\be
[A(\g)\stackrel{\otimes}{,}A(\mu)]
= i\hbar \Big[r\;,\;A(\g)\otimes \I + \I\otimes A(\mu)\Big]
\la{CR}\ee
\baa
{}[ \x, (\log h)_\x  ] &=& [\xb, (\log h)_{\xb} ] ~=~ i\hbar
\label{CR1}\\ 
{}[ \xb, (\log h)_\x  ] &=& [\x, (\log h)_\xb ] ~=~ 0\nn
\eaa
Suppose now that all $\g_j$ are imaginary (i.e. $w_j\in \R$); then
by Thm.\r{constr} we should require all elements of $A_j$ 
to be real at the classical level. Quantum mechanically we get
\be
A_j\equiv  \f{i\hbar}{2}
   \pmatrix{ h_j & 2e_j \cr 2f_j & -h_j \cr}, 
\la{Aj}\ee
where $e_j, f_j, h_j$ are the anti-hermitian Chevalley generators of
$SL(2,\R)$ obeying the standard commutation relations
\baa
{}[h_j,\;e_j] &=& 2e_j\la{CR2}\\
{}[h_j,\;f_j] &=& -2f_j\non
{}[e_j,\;f_j] &=& h_j\nn
\eaa
Unitary representations of \Ref{CR2} with Casimir operator
\ben
\f{-4}{\hbar^2}\tr A_j^2= \f{1}{2}h_j^2 +e_j f_j + f_j e_j
\een
equal to
$s_j(s_j-2)$ are given by
\baa
e_j &=&  \z_j^2 \f{d}{d\z_j}+ s_j\z_j\la{gener}\\
f_j &=&-\f{d}{d\z_j}\non
h_j &=&2\z_j \f{d}{d\z_j}+s_j,\nn
\eaa
where $\{ \z_j \}$ are the arguments of the  functions spanning the
representation space $\H_j$, which  may belong to the
principal, supplementary or discrete series of $SL(2,\R)$.

According to \Ref{CR1} one can choose
\be
(\log h)_{\x} = - i\hbar \f{\p}{\p \x} \;\;\;\;\;\;
(\log h)_{\xb} =- i\hbar \f{\p}{\p \xb}
\la{hq} \ee
Thus the wave function $\Phi$ of a given isomonodromic sector with
$w_j\in\R$ should depend on $(\x,\xb)$ and live in the
direct product
\ben
\H^{(N)}=\H_1\otimes\dots\otimes \H_N
\een
of $N$ unitary representation spaces of $SL(2,\R)$.
This means that $\Phi$ can be realized as a function 
\ben
\Phi\equiv\Phi(\x,\xb ;\z_1,\dots , \z_N).
\een

\section{Wheeler-DeWitt equations and Knizh\-nik-Zamolodchikov
system for $SL(2,\R)$} 

The Wheeler-DeWitt equations now take the form
\be
\f{d\Phi}{d\x}=\f{d\Phi}{d\xb}=0
\ee
or, equivalently,
\be
\Cx\Phi= \Cxb\Phi = 0
\ee
which can be written out by use of the explicit form of the
constraints $\Cx$ and $\Cxb$ given in \Ref{Ham}
\Ref{Aj},\Ref{gener} and \Ref{hq}:
\baa
\f{\p\Phi}{\p\x}&=& 
-i\hbar \sum_{k \neq j} \f{\Omega_{jk}}{(1-\g_j)(1-\g_k)}\Phi\la{WDW2}\\
\f{\p\Phi}{\p\xb}&=& 
-i\hbar \sum_{k\neq j} \f{\Omega_{jk}}{(1+\g_j)(1+\g_k)}\Phi\nn
\eaa
where 
\baa
\Omega_{jk}&\equiv& \f{1}{2}h_jh_k+ e_j f_k +e_k f_j \non
&=& -(\z_j -\z_k)^2 \f{\p^2}{\p\z_j\z_k} \la{WDWG}\\
&&{} +(\z_k-\z_j)\big(s_j\f{\p}{\p\z_k}-
s_k\f{\p}{\p\z_j}\big) + \f{s_j s_k}{2} \nn
\eaa

According to Thm.\ref{constr}, the wave functionals satisfying the
coset constraints should be symmetric with respect to the
involution $\g_j\rightarrow 1/\g_j$ and satisfy the constraint
\be
\sum_{j=1}^N A_j\Phi = 0 .
\la{af}\ee

The general solution of the system \Ref{WDW2} is not known. 
However, these equations turn out to be intimately related to the
Knizhnik-Zamolodchikov system for $SL(2,\R)$ \cite{KZ,Liouville}:
\be
\f{\p \Phi_{KZ}}{\p\g_j}=-i\hbar\sum_{k\neq j}\f{\Omega_{jk}}
{\g_j-\g_k} \Phi_{KZ}
\la{KZ}\ee
with an ${\H}^{(N)}$-valued function $\Phi_{KZ}(\x,\xb)$. 

\begin{Theorem}
If $\Phi_{KZ}$ is annihilated by the ``total spin"
\ben
\sum_{j=1}^N A_j \Phi_{KZ}=0
\een
and the $\g_j$ depend on $(\x,\xb)$ according to \Ref{gamma}, 
then 
\be
\Phi = 
\prod_{j=1}^{N} \bigg( \f{\partial \g_j}{\partial w_j}
      \bigg)^{-\f{1}{4}\hbar^2 s_j (s_j-2)} 
 \Phi_{KZ} 
\la{link5}\ee
solves the constraint (Wheeler DeWitt) equations \Ref{WDW2}.
\end{Theorem}
Thus, the task of solving \Ref{WDW2} reduces to the solution of
\Ref{KZ}.

The full set of quantum observables is related to the algebra of
monodromies for the KZ equations \Ref{KZ} which is well understood
only for $SU(2)$ where it gives rise to certain quantum groups
\c{Drinfeld}.

The only solutions of KZ equations for the non-compact group $SL(2,\R)$ 
known so far are solutions corresponding to the 
unitary discrete series representations (either positive
for all $j$ or negative for all $j$) all of which possess a ground 
(lowest weight) state in $\H^{(N)}$.  However, it is
possible to show \c{KN3} that solutions of this kind cannot satisfy
the constraint \Ref{af}.
Moreover, a simple analysis of the sign of the  Casimir operator shows
that on order to construct wave functions corresponding to physically
interesting classical solutions (such as Kerr-NUT) one would have to
consider representations of the continuous series.
Namely, for all known classical solutions (including Kerr-NUT) we
have $\tr A_j^2 > 0$. However, in the quantum case,
\ben
\tr A_j^2= -\f{\hbar^2}{2} s(s-2)
\een
For the discrete series, when $s$ is real and integer, this is always
negative. For the continuous series we have $s=1+iq$ with $q\in\R$, and
the eigenvalue of $\tr A_j^2$ is positive.

In the next section we shall briefly discuss how one might go about
constructing a quantum state whose semiclassical limit would
reduce to the Kerr-NUT solution.

\section{Towards a quantum Kerr solution}
According to the previous section, the desired solution of 
\Ref{KZ} for $N=4$ is a function of the four positions of the poles
$\g_i$ defined by \Ref{wKerr} and of the four auxiliary variables
$\z_i$, on which the algebra $sl(2,{\bf R})$ is represented.

The constraints \Ref{af} have to annihilate this function which
hence is $SL(2,{\R})$ invariant. Therefore it essentially depends
only on the ratio
\ben
\frac{(\z_1-\z_2)(\z_3-\z_4)}{(\z_1-\z_4)(\z_3-\z_2)}
\een
just as in conformal field theory where the conformal Ward identities
reduce the correlation functions to a function of a single
variable \cite{BPZ}. Moreover, the validity of
the KZ equations implies an analogous reduction of the
$\g_i$-dependence.

The quantum state then reduces to the following form:
\be
\Phi = 
\prod_{j=1}^{4} \bigg( \f{\partial \g_j}{\partial w_j}
      \bigg)^{-\f{1}{4}\hbar^2 s (s-2)} F(\z_i,\g_i), 
\ee
with
\baa
s &=& s_1=s_2=s_3=s_4\non
F(\z_i,\g_i) &=& 
  (\g_1-\g_4)^{-\D}(\g_2-\g_3)^{-\D}\times\non
&&(\z_1-\z_4)^{-s}(\z_2-\z_3)^{-s} G(x,y),\nn
\eaa
\baa
x &=& \frac{(\z_1-\z_2)(\z_3-\z_4)}{(\z_1-\z_4)(\z_3-\z_2)}\non
y &=& \frac{(\g_1-\g_2)(\g_3-\g_4)}{(\g_1-\g_4)(\g_3-\g_2)}\non
{}\D &=& \frac{i\hbar}2 s(s-2)\nn
\eaa 

The remaining KZ equation for the function $G$ 
can be obtained by a lengthy but straightforward calculation
which gives
\be
\partial_y G(x,y) = 
i\hbar\Big(\frac{D(x)}{y}-\frac{D(1-x)}{1-y}\Big)G(x,y)
\la{qKerr}
\ee
with
\baa
D(x) &=& x^2(1-x)\partial^2_x+2sx(1-x)\partial_x\non
&&{} +\frac12s^2(1-2x)\nn
\eaa

An equivalent form of this equation appeared in the study of four-point 
correlation-functions in Liouville theory \cite{FZ}.

Equation \Ref{qKerr} is very similar to the standard hypergeometric
equation, where $D(x)$ and $D(1-x)$ are just two $2\times 2$
matrices.  The singular points $y= 0, 1,\i$ have a
very definite physical meaning from the point of view of the classical
Kerr solution.  Namely, we can express the variable $y$ in terms of
prolate ellipsoidal coordinates \Ref{XY} as follows:
\ben
y=\f{1-\Y^2}{1-\X^2}=-\f{\rho^2}{\s^2(\X^2-1)^2}
\een
This shows that $y=0$ corresponds classically to the spatial infinity
and the part of the symmetry axis outside of the event horizon.
The value $y=1$ corresponds to the poles of the event horizon, and
$y=\i$ corresponds to the surface of the event horizon.
 
The analysis of equation \Ref{qKerr} should give asymptotical
expansions of the wave functional at these singular points and
allow us to relate them.  This would then enable us
to understand the behavior of physically interesting
expectation values at these points and to clarify the meaning and
the fate of the classical singularities in the quantum theory.
  
The classical limit leading to the Kerr-NUT solution should look like
\ben
\f{\hbar^2}{4} (1+ q^2) \rightarrow 1
\een
If this limit is equal to an integer $k$, the related classical solution
should be the $k$th member of the Tomimatsu-Sato hierarchy.

\section*{Acknowledgments} The work of D.~Korotkin was supported by DFG
contract Ni 290/5-1; H.~Samtleben was supported by Studienstiftung des
deutschen Volkes.

\end{document}